\documentclass[11pt]{article}
\usepackage{longtable}
\usepackage{graphicx}
\usepackage{lineno}
\usepackage{color}
\definecolor{darkgreen}{rgb}{0,0.5,0}
\usepackage{hyperref}
\hypersetup{colorlinks=true, urlcolor=blue, citecolor=darkgreen}
\usepackage{natbib}
\usepackage{fullpage}
\usepackage{setspace}
\usepackage{listings}

\begin{document}
\begin{flushleft}

{\Large
\textbf{Improving transcriptome assembly through error correction of high-throughput sequence reads}
}
\\ 
\vspace{4mm}

\noindent
Matthew D MacManes$^{1}$$^\ast$ and
Michael B. Eisen$^{1,2}$ \\
\vspace{5mm}

\bf{1} \textnormal{UC Berkeley. California Institute of Quantitative Biology, Berkeley, CA, USA} \\
\bf{2} \textnormal{Howard Hughes Medical Institute} \\
\vspace{2mm}
 
\bf{$\ast$} \textnormal{Corresponding author: \href{mailto:macmanes@gmail.com}{macmanes@gmail.com}, Twitter: \href{https://twitter.com/PeroMHC}{$@$PeroMHC}}
\end{flushleft}
\vspace{4mm}

\begin{abstract}
\noindent
The study of functional genomics--particularly in non-model organisms has been dramatically improved over the last few years by use of transcriptomes and RNAseq. While these studies are potentially extremely powerful, a computationally intensive procedure--the \textit{de novo} construction of a reference transcriptome must be completed as a prerequisite to further analyses. The accurate reference is critically important as all downstream steps, including estimating transcript abundance are critically dependent on the construction of an accurate reference. Though a substantial amount of research has been done on assembly, only recently have the pre-assembly procedures been studied in detail. Specifically, several stand-alone error correction modules have been reported on, and while they have shown to be effective in reducing errors at the level of sequencing reads, how error correction impacts assembly accuracy is largely unknown. Here, we show via use of a simulated and empiric dataset, that applying error correction to sequencing reads has significant positive effects on assembly accuracy, and should be applied to all datasets.  A list of commands with will allow for the production of \textsc{Reptile} corrected reads is available at \url{https://gist.github.com/macmanes/5878728}
\end{abstract}

\linenumbers

\section*{Introduction}

\doublespacing
\noindent
The popularity of genome enabled biology has increased dramatically, particularly for researchers studying non-model organisms, during the last few years.  For many, the primary goal of these works is to better understand the genomic underpinnings of adaptive \citep{Linnen:2013ir,Narum:2013kc} or functional \citep{MunozMerida:2013dw,Hsu:2012dg} traits. While extremely promising, the study of functional genomics in non-model organisms typically requires the generation of a reference transcriptome to which comparisons are made.  Although compared to genome assembly \citep{Bradnam:2013uu,Earl:2011gt}, transcriptome assembly is less challenging, significant hurdles still exist (see \cite{Francis:2013gc,Vijay:2012gy,Pyrkosz:2013tm} for examples of the types of challenges). \\

\noindent
The process of transcriptome assembly is further complicated by the error-prone nature of high-throughput sequencing reads.  With regards to Illumina sequencing, error is distributed non-randomly over the length of the read, with the rate of error increasing from 5' to 3' end \citep{Liu:2012iv}. These errors are overwhelmingly substitution errors \citep{Yang:2013ck}, with the global error rate being between 1\% and 3\%.    While beyond the focus of this paper, the accuracy of \textit{de novo} transcriptome assembly, sequencing errors may have important implications for SNP calling, and the estimation of nucleotide polymorphism and the estimation of transcript abundance. \\

\noindent
With regards to assembly, sequencing read error has both technical and 'real-world' importance. Because most transcriptome assemblers use a \textit{de Bruijn} graph representation of sequence connectedness, sequencing error can dramatically increase the size and complexity of the graph, and thus increase both RAM requirements and runtime \citep{Conway:2011iv,Pell:2012id}. More important, however, are their effects on assembly accuracy. Before the current work, sequence assemblers were thought to efficiently handle error given sufficient sequence coverage. While this is largely true, sequence error may lead to assembly error at the nucleotide level despite high coverage, and therefore should be corrected, if possible. In addition, there may be technical, biological, or financial reasons why extremely deep coverage may not be possible, therefore, a more general solution is warranted.  \\ 

\noindent
While the vast majority of computational genomics research has focused on either assembly \citep{Chaisson:2004kf,Miller:2010gxa,Earl:2011gt,Bradnam:2013uu} or transcript abundance estimation \citep{Soneson:2013fr,Marioni:2008bg,Mortazavi:2008jj,Pyrkosz:2013tm}, up until recently, research regarding the dynamics of pre-assembly procedures has largely been missing. However, error correction has become more popular, with several software packages becoming available for error correction, e.g.  \textsc{AllPathsLG} error correction \citep{Gnerre:2011fd}, \textsc{Quake} \citep{Kelley:2010kg}, \textsc{Echo} \citep{Kao:2011bx}, \textsc{Reptile} \citep{Yang:2010kv}, SOAP\textit{denovo} \citep{Liu:2011ez}, \textsc{SGA} \citep{Simpson:2010fd} and \textsc{Seecer} \citep{Le:2013dy}.  While these packages have largely focused on the error correction of genomic reads (with exception to \textsc{Seecer}, which was designed for RNAseq reads), they may likely be used as effectively for RNAseq reads.  \\ 

\noindent
Recently a review \citep{Yang:2013ck} evaluating several of these methods in their ability to correct genomic sequence read error was published. However, the application of these techniques to RNAseq reads, as well as an understanding of how error correction influences accuracy of the \textit{de novo} transcriptome assembly has not been evaluated. Here we aim to evaluate several of the available error correction methods. Though an understanding of the error correction process itself, including it's interaction with coverage may be a useful exercise, our initial efforts described here, focus on the the effects of error correction on assembly, the resource which forms the basis of all downstream (e.g. differential expression, SNP calling) steps.  \\

\noindent
To accomplish this, we simulated 30 million paired-end Illumina reads and assembled uncorrected reads, as well as reads corrected by each of the evaluated correction methods, which were chosen to represent the breadth of computational techniques used for sequence read error correction.  Though we focus on the simulated dataset, we corroborate our findings through use of an empirically derived Illumina dataset.  For both datasets, we evaluate assembly content, number of errors incorporated into the assembly, and mapping efficiency in an attempt to understand the effects of error correction on assembly. Although Illumina is just one of the available high-throughput sequencing technologies currently available, we chose to limit our investigation to this single, most widely used technology, though similar investigations will become necessary as the sequencing technology evolves. \\

\noindent
Because the \textit{de novo} assembly is a key resource for all subsequent studies of gene expression and allelic variation, the production of an error-free reference is absolutely critical. Indeed, error in the reference itself will have potential impacts on the results of downstream analyses. These types of error may be particularly problematic in \textit{de novo} assemblies of non-model organisms, where experimental validation of sequence accuracy may be impossible. Though methods for the correction of sequencing reads have been available for the last few years, their adoption has been limited, seemingly because a demonstration of their effects has been lacking. Here, we show that error correction has a large effect on assembly quality, and therefore argue that it should become a routine part of workflow involved in processing Illumina mRNA sequence data. Though this initial work focuses on the results of error correction; arguably the most logical candidate for study, future work will attempt to gain a deeper understanding of error in the error correction process itself. \\

\section*{Results}
\noindent
Thirty million 100nt paired-end (PE) reads were simulated using the program \textsc{Flux Simulator} \citep{Griebel:2012ti}.  Simulated reads were based on the coding portion of the \textit{Mus musculus} genome and included coverage of about 60k transcripts with average depth of 70X. Thirty million reads were simulated as this corresponds to the sequencing effort suggested by \citep{Francis:2013gc} as an appropriate effort, balancing coverage with the accumulation of errors, particularly in non-model animal transcriptomics.  These reads were qualitatively similar to several published datasets \citep{MacManes:2012bu,Chen:2011ba}. Sequence error was simulated to follow the well-characterized Illumina error profile (Supplementary Figure 1).  Similarly, patterns of gene expressions were typical of many mammalian tissues (Supplementary Figure 2), and follows a Poisson distribution with lambda=1 \citep{Auer:2011bt,Hu:2011cn,Jiang:2009bw}.  \\

\noindent
In addition to the simulated dataset, error correction was applied to an empirically derived Illumina dataset. This dataset consists of 50 million 76nt paired-end Illumina sequence reads from \textit{Mus musculus} mRNA, and is available as part of the Trinity software package \citep{Haas:jq,Grabherr:2011jb}. Because we were interested in comparing the two datasets, we randomly selected 30 million PE reads from the total 50 million reads for analyses.  The simulated read dataset is available at \url{https://www.dropbox.com/s/mp8fu0tijox69ki/simulated.reads.tar.gz}, while the empirical dataset is at \url{https://www.dropbox.com/s/rkl0ihqom28smb2/empiric.reads.tar.gz}.  [Of note, these datasets are to be moved to Dryad upon acceptance for publication. ]\\

\noindent
Error correction of the simulated and empiric datasets was completed using the \textsc{Seecer}, \textsc{AllPathsLG}, \textsc{SGA}, and \textsc{Reptile} error correction modules.  Details regarding the specific numbers of nucleotide changes and the proportion of reads being affected are detailed in (\hyperlink{Table 1}{Table 1}). Despite the fact that each software package attempted to solve the same basic problem, runtime considerations and results were quite different. \textsc{Trinity} assembly using the uncorrected simulated reads produced an assembly consisting of 78.43Mb, while the assembly of empirically derived reads was 74.24Mb.  \\

\noindent
\textbf{\hypertarget{Table 1}{Table 1}} \\
\begin{center}
    \begin{tabular}{ | l | l | l  | l | l |}
    \hline
    Simulated Dataset &  Total Reads & Num reads corr & Num  nt corr & Runtime  \\ \hline
    Raw reads & 30M PE & n/a & n/a & n/a \\ \hline
    \textsc{AllPathsLG} Corr. & 30M PE & ? & 139,592,317 & \ $\sim$ 8hrs \\ \hline
    \textsc{Sga} Corr. & 30M PE & ? & 19,826,919 & \ $\sim$ 38 minutes \\ \hline
    \textsc{Reptile} Corr. & 30M PE & 2,047,088 & 7,782,594 & \ $\sim$ 3 hours \\ \hline
     \textsc{Seecer} Corr. & 30M PE & 8,782,350 & 14,033,709 & \ $\sim$ 5 hours \\ \hline
       \end{tabular}
\\
\end{center}
\noindent
\begin{changemargin}{2cm}{2cm}
Table 1. Number of raw sequencing reads, sequencing reads corrected, nucleotides (nt) corrected, and approximate runtime for each of the datasets. Note that neither \textsc{AllPaths} nor \textsc{Sga} provides information regarding the number of reads affected by the correction process. 
\end{changemargin}

\subsection*{Simulated Data}
Analyses focused on a high-confidence subset of the data, as defined as being 99\% similar to the reference over at least 90\% of its length. The high-confidence subset of the simulated uncorrected read assembly (n=38459 contigs) contained approximately 54k nucleotide mismatches (\hyperlink{Figure 1}{Figure 1}), corresponding to an mean error rate of 1.40 mismatches per contig (SD=7.38, max=178).  There did not appear to be an observe an obvious relationship between gene expression and the quality of the assembled transcripts (\hyperlink{Figure 2}{Figure 2}). While the rate of error is low, and indeed a testament to the general utility the \textit{de Bruijn} graph approach for sequence assembly, a dramatic improvement in accuracy would be worth pursuing, if possible.  \\

\noindent
Error correction of simulated reads using \textsc{Reptile} was a laborious process, with multiple (\textgreater 5) individual executions of the program required for parameter optimization. While each individual run was relatively quick, the total time exceed 12 hours, with manual intervention and decision making required at each execution.  Error correction resulted in the correction of 7.8M nucleotides (of a total $\sim$ 5B nucleotides contained in the sequencing read dataset). The resultant assembly contains an average of 1.23 mismatches per contig (SD=6.46, max=152).  The absolute number of errors decreased by $\sim$ 12\% (\hyperlink{Figure 1}{Figure 1}), which represents substantial improvement, particularly given that the high confidence subset of the Reptile-corrected assembly was the largest (n= 38670 contigs) of any of the methods (\hyperlink{Table 2}{Table 2}). \\

\textbf{\hypertarget{Figure 1}{Figure 1}} \\
\centerline{\includegraphics[width=12.0\baselineskip]{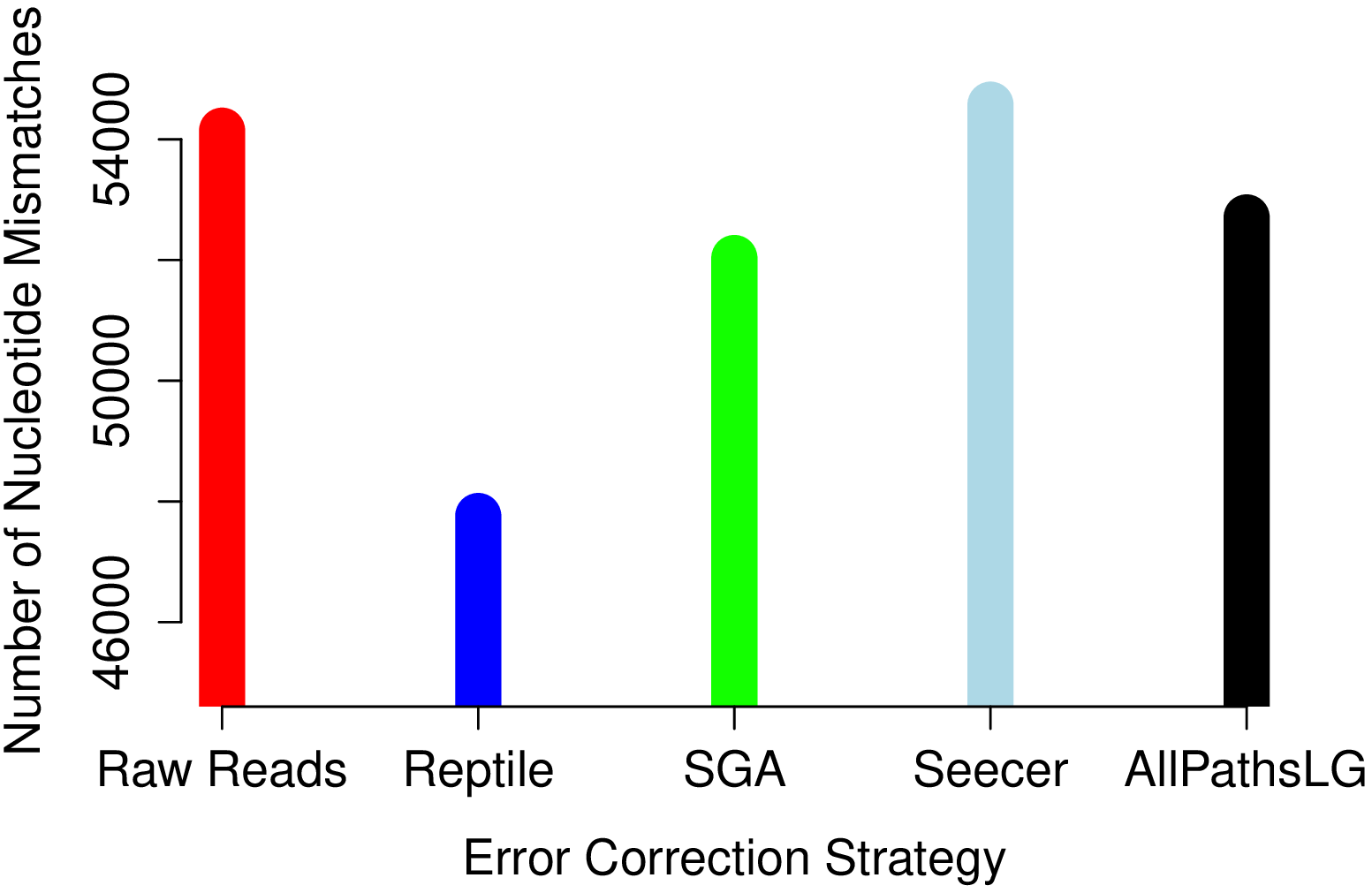}}
\begin{changemargin}{2cm}{2cm}
\noindent
Fig. 1. The global estimate of nucleotide mismatch decreases with error correction. The assembly done with \textsc{Reptile} corrected reads has approximately 10\% fewer errors than does the raw read assembly. 
\end{changemargin}
\vspace{10mm}

\noindent
\textsc{AllPathsLG} error correction software implemented by far the most aggressive correction, selected optimized parameters in an automated fashion, and did so within a 4 hour runtime.  \textsc{AllPathsLG} corrected nearly 140M nucleotides (again, out of a total $\sim$ 5B nucleotides contained in the sequencing reads), which resulted in a final assembly with 52706 nucleotide errors, corresponding to a decrease in error of approximately 2.7\%.   \\

\noindent
\textsc{Seecer}, is the only dedicated error-correction software package dedicated to RNAseq reads. Though \textsc{Seecer} is expected to handle RNAseq datasets better than the other correction programs, its results were disappointing. More than 14 million nucleotides were changed, affecting approximately 8.8M sequencing reads.  Upon assembly 54,574 nucleotide errors remained, which is equivalent to the number of errors contained in the assembly of uncorrected reads.   \\

\textbf{\hypertarget{Figure 2}{Figure 2}} \\
\centerline{\includegraphics[width=22.0\baselineskip]{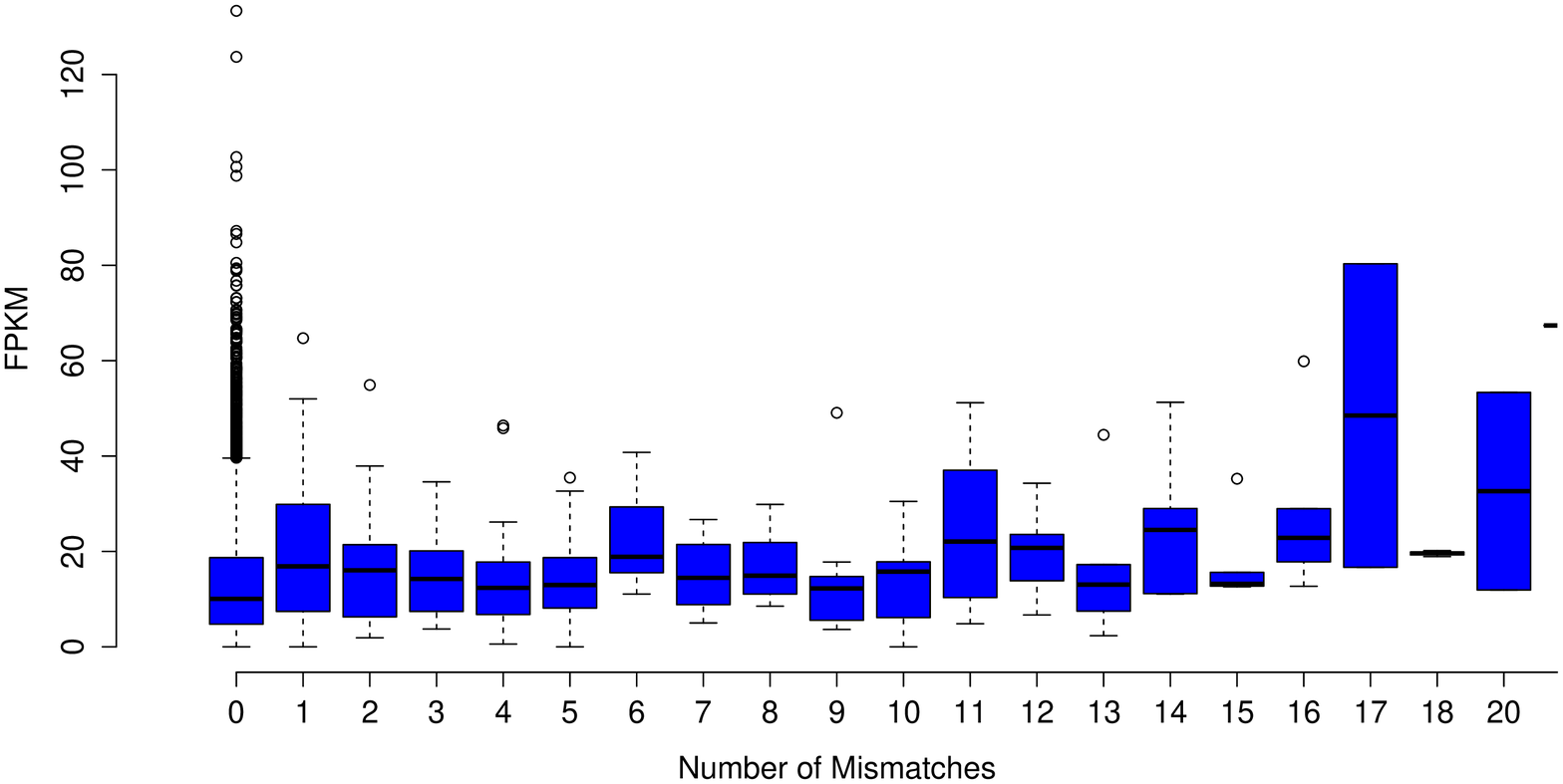}}
\begin{changemargin}{2cm}{2cm}
\noindent
Fig. 2. The number of nucleotide mismatches in a given contig is not related to gene expression. On average, in the assembly of uncorrected simulated reads, poorly expressed transcripts are no more error prone than are highly expressed transcripts.
\end{changemargin}
\vspace{10mm}

\noindent 
Lastly, \textsc{Sga} error correction was implemented on the simulated read dataset. \textsc{Sga}, is the fastest of all error correction modules, and finished correcting the simulated dataset in 38 minutes.  The software applied corrections to 19.8M nucleotides.  It's correction resulted in a modest improvement in error, with a reduction in error of approximately 4\% over the assembly of uncorrected errors.  \\
 
\noindent
Assembly content, aside from fine--scaled differences at the nucleotide level, as described above, were equivalent. Assemblies consisted of between 63,099 (\textsc{Reptile}) -- 65,468 (\textsc{Seecer}) putative transcripts greater than 200nt in length. N50 ranged from 2319 (\textsc{Reptile}) -- 2403nt (\textsc{Sga}).  The high-confidence portion of the assemblies ranged in size from 38407 contigs (\textsc{Seecer} assembly) to 38670 contigs in the \textsc{Reptile} assembly. Assemblies are detailed in \hyperlink{Table 2}{Table 2}, and available at \url{http://dx.doi.org/10.6084/m9.figshare.725715}. \\

\noindent
\textbf{\hypertarget{Table 2}{Table 2}} \\
\begin{center}
    \begin{tabular}{ | l | l | l  | l | l |}
    \hline
    Dataset & Error Corr. Method &  Raw Assembly Size & High Conf. Size  \\ \hline
    Simulated Reads & & & \\ \hline
     & None & 64491 (78Mb)&38459 (27Mb)  \\ \hline
     & \textsc{AllPathsLG} &64682 (78Mb) & 38628 (27Mb) \\ \hline
     & \textsc{Sga} &65059 (80Mb) & 38619 (27Mb)\\ \hline
     & \textsc{Reptile} &63099 (73Mb) &38670 (25Mb)  \\ \hline
     & \textsc{Seecer} &65468 (80Mb) & 38407 (27Mb) \\ \hline
Empiric Reads & & & \\ \hline
     & None &57338 (74Mb) & 21406 (24Mb)\\ \hline
     & \textsc{AllPathsLG} &53884 (66Mb) & 21204 (23Mb)\\ \hline
     & \textsc{Sga} &56707 (75Mb) & 21323 (24Mb) \\ \hline
     & \textsc{Reptile} &53780 (60Mb) & 21850 (22Mb)\\ \hline
     & \textsc{Seecer} &57311 (75Mb) & 21268 (24Mb) \\ \hline

  \end{tabular}
\\
\end{center}
\noindent
\begin{changemargin}{2cm}{2cm}
Table 2. Assembly details. High confidence datasets included only contigs that matched a single reference, had sequence similarity \textgreater 99\%, and covered $\geq$ 90\% of length of reference. 
\end{changemargin}

\vspace{10mm}
\noindent
The proportion of reads mapping to each assembled dataset was equivalent as well, ranging from 92.44\% using raw reads to 94.89\% in \textsc{Sga} corrected reads.  Assemblies did not appear to differ in general patterns of contiguity,  (\hyperlink{Figure 3}{Figure 3}), though it should be noted that the most successful error corrector, \textsc{Reptile} had both the smallest assembly size \textit{and} largest number of high confidence contigs.  Taken together, these patterns suggest that error correction may have a significant effect on the structure of assembly; though its major effects are in enhancing resolution at the level of the nucleotide. Indeed, while we did not find, nor expect to find large differences in these global metrics, we do expect to see a significant effect on transcriptome based studies of marker development and population genetics, which are endeavors fundamentally linked to polymorphism, estimates of which can easily be confused by sequence error.    \\
 
\subsection*{Empirical Data} 
The high-confidence subset of the uncorrected empirical read assembly (n=21406 contigs) contained approximately 14.7k nucleotide mismatches, corresponding to an mean error rate of .68 mismatches per contig (SD=3.60 max=197).  Error correction procedures were implemented as described above. Indeed, the resultant pattens of correction were recapitulated. Error correction using \textsc{Reptile} were most favorable, and resulted in a reduction in the number of nucleotide errors by more than 10\%, to approximately 13k.  As above, the high-confidence portion of the \textsc{Reptile}-corrected dataset was the largest, with 21580 contigs, which is slightly larger that the assembly of uncorrected reads. Similar to what was observed in the simulated dataset, the high-confidence portion of the \textsc{AllPaths} corrected assembly was the smallest of any of the datasets, and contained the most error. Of interest, the \textsc{Sga} correction performed well, similar to as in simulated reads, decreasing error by more than 9\%. \\

\noindent
Empirical assemblies contained between 53780 (\textsc{Reptile}) and 57338 (uncorrected assembly) contigs greater than 200nt in length.  N50 ranged from between 2412 (\textsc{Reptile}) and 2666nt (\textsc{Seecer}) in length.  As above, assemblies did not differ widely in their general content or structure; instead effects were limited to differences at nucleotide level. Assemblies are available at \url{http://dx.doi.org/10.6084/m9.figshare.725715}. \\

\textbf{\hypertarget{Figure 3}{Figure 3}} \\
\centerline{\includegraphics[width=20.0\baselineskip]{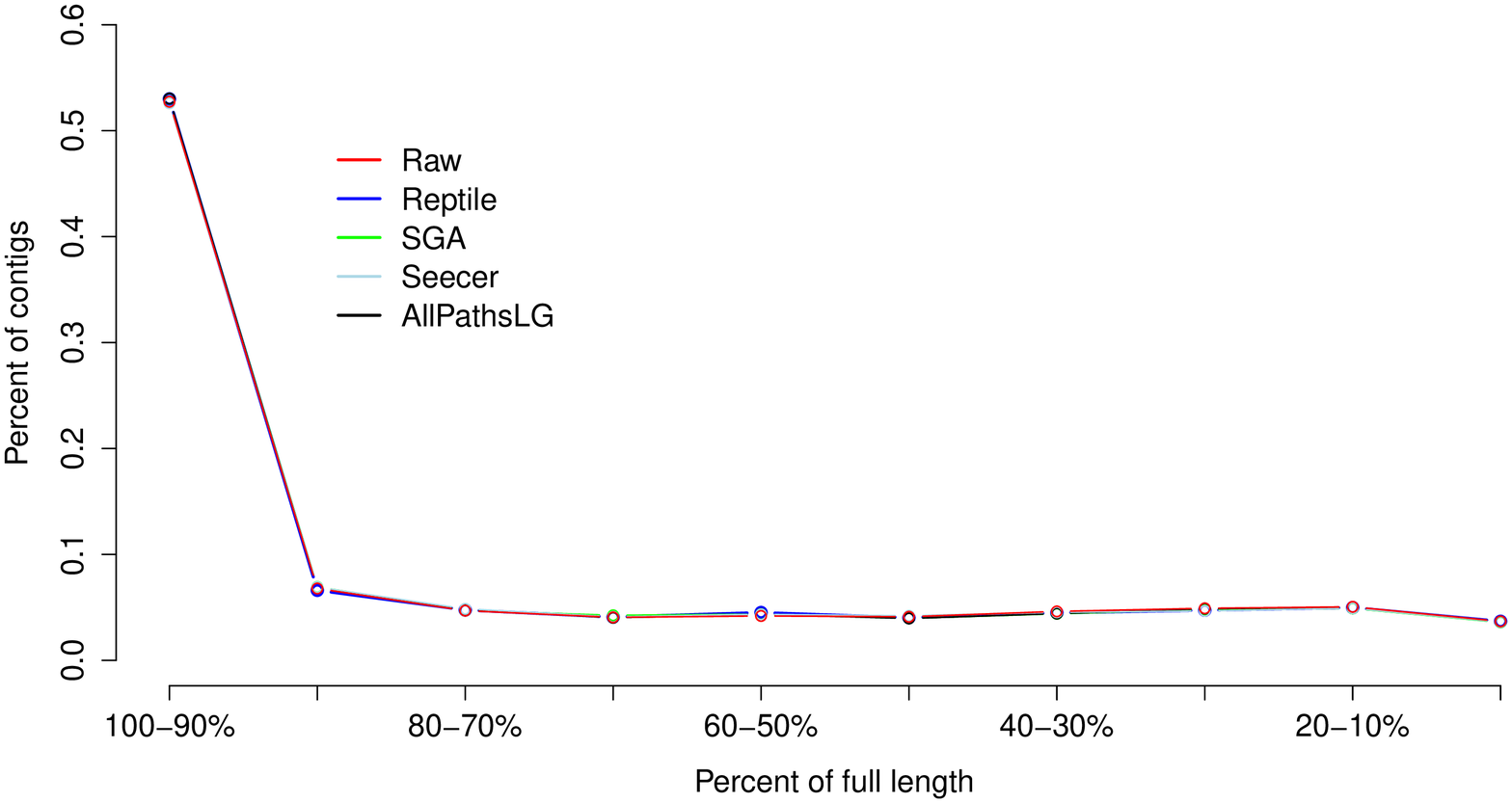}}

\noindent
\begin{changemargin}{2cm}{2cm}
Fig. 3. Assembly contiguity did not vary significantly between assemblies of reads using the different error correction methods. Each error correction methods, as well as assembly of raw reads, produced an assembly that is dominated by full length (both start and stop codon present) or nearly full length assembled transcripts. 
\end{changemargin}
\vspace{10mm}

\section*{Discussion}

Though the methods for error correction have become increasingly popular within the last few years, their adoption in general genome or transcriptome assembly pipelines has lagged. One potential reason for this lag has been that their effects on assembly, particularly in RNAseq, has not been demonstrated. Here, we attempt to evaluate the effects of four different error correction algorithms on assembly- arguably the step upon which all downstream steps (e.g. differential expression, functional genomics, SNP discovery, etc.) is based. We use both simulated and empirically derived data to show a significant effect of correction on assembly-- especially when using the error corrector \textsc{Reptile}. This particular method, while relatively labor intensive to implement, reduces error by more than 10\%, and results in a larger high-confidence subset relative to other methods.  Aside from a reduction in the total number of errors, \textsc{Reptile} correction both reduced variation in nucleotide error, and reduced the maximum number of errors in a single contig.   \\

\noindent
Interesting, \textsc{Seecer}, the only error correction method designed for RNAseq reads, performed relatively poorly. In simulated reads, \textsc{Seecer} slightly increased the number of errors in the assembly, though with applied to empirically derived reads, results were more favorable, decreasing error by $\sim$ 3\%.  Though the effects of coverage on correction efficiency were not explored in the manuscript describing \textsc{Seecer} \citep{Le:2013dy}, their empirical dataset contained nearly 90 million sequencing reads, a size 3X larger than the dataset we analyze here.  Future work investigating the effects of coverage on error correction is necessary.  \\

\noindent
In addition to this, how error correction interacts with the more complicated reconstructions, splice variants for instance, is an outstanding question. Indeed, reads traversing a splicing junction may be particularly problematic for error correctors, as coverage on opposite sides of the junction may be different owing to differences in isoform expression, which could masquerade as error.  Alternative splicing is known to negatively affect both assembly and mapping \citep{Vijay:2012gy,Sammeth:2009jx,Pyrkosz:2013tm}, and given that many computational strategies are shared between these techniques and error correction suggests that similarly, error correction should be affected by splicing. Indeed, many of the most error-rich contigs were those where multiple isoforms were present.  As such, considering this potential source of error in error correction should be considered during error correction. Computational strategies that distinguish these alternative splicing events from real error are currently being developed.   \\
 
\noindent
The effects of read coverage on the efficiency of error correction are likely strong.  Aside from the suggestion that \textsc{Seecer's} relatively poor performance owed to low coverage data relative to the dataset tested during the development of that software \citep{Le:2013dy}, other supporting evidence exists.  Approximately 5\% of reads are miscorrected.  When looking at a sample (n=50000) of these reads, the contig to which that read maps is on average more lowly expressed than appropriately corrected reads (\hyperlink{Figure 4}{Figure 4}, Wilcoxon rank sum test, W = 574733, p-value = 0.00022), which suggests that low coverage may reduce the efficiency of error correction. In addition, miscorrected reads, whose average expression is lower, tend to have more corrections than to the appropriately corrected reads (\hyperlink{Figure 5}{Figure 5}, t test, t = -2.1755, df = 7164.8, p-value = 0.029).

\textbf{\hypertarget{Figure 4}{Figure 4}} \\
\centerline{\includegraphics[width=20.0\baselineskip]{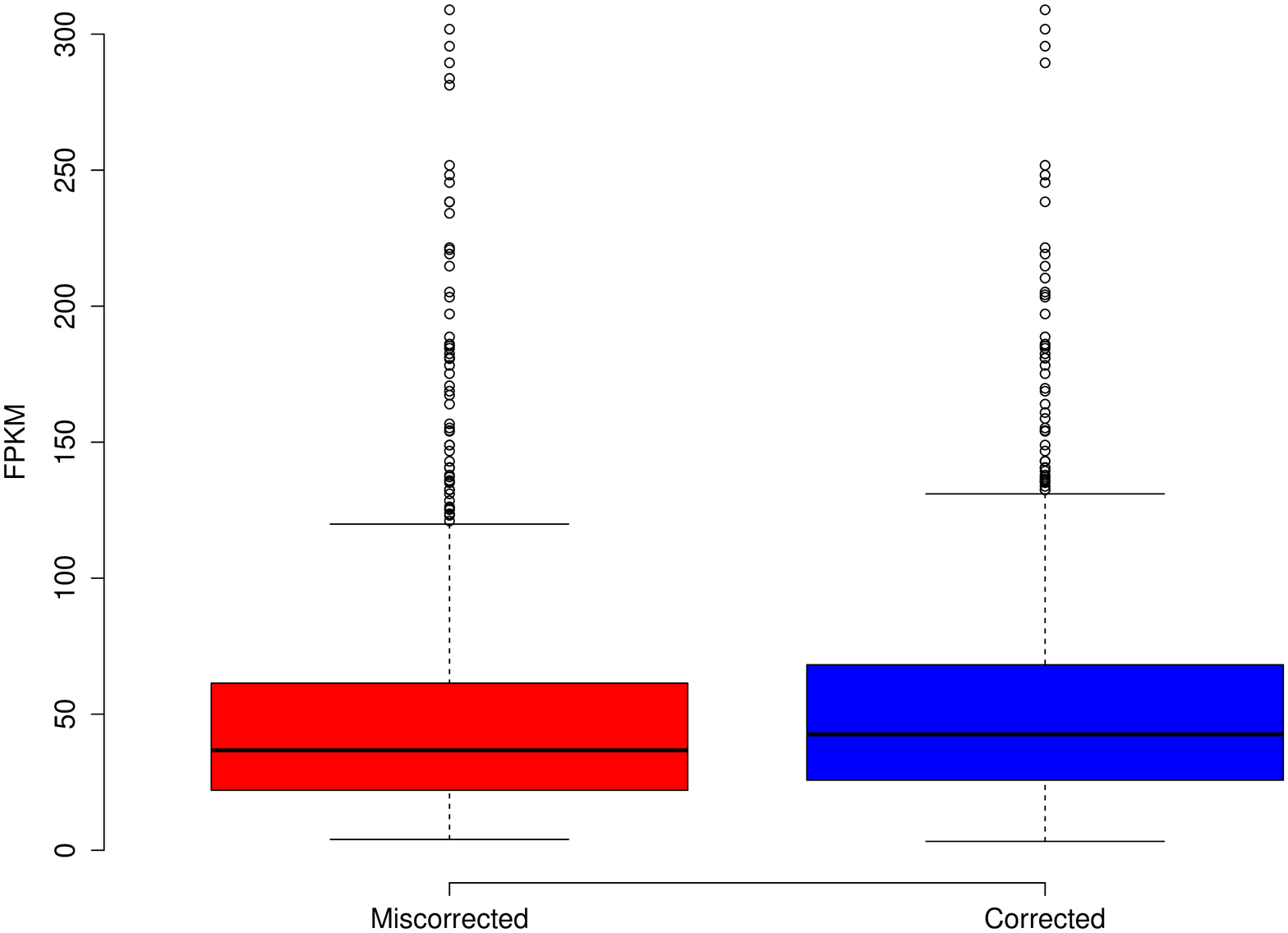}}

\noindent
\begin{changemargin}{2cm}{2cm}
Fig. 4. Reads miscorrected by \textsc{Reptile} have lower expression, on average, than to appropriately corrected reads.
\end{changemargin}
\vspace{10mm}

\textbf{\hypertarget{Figure 5}{Figure 5}} \\
\centerline{\includegraphics[width=20.0\baselineskip]{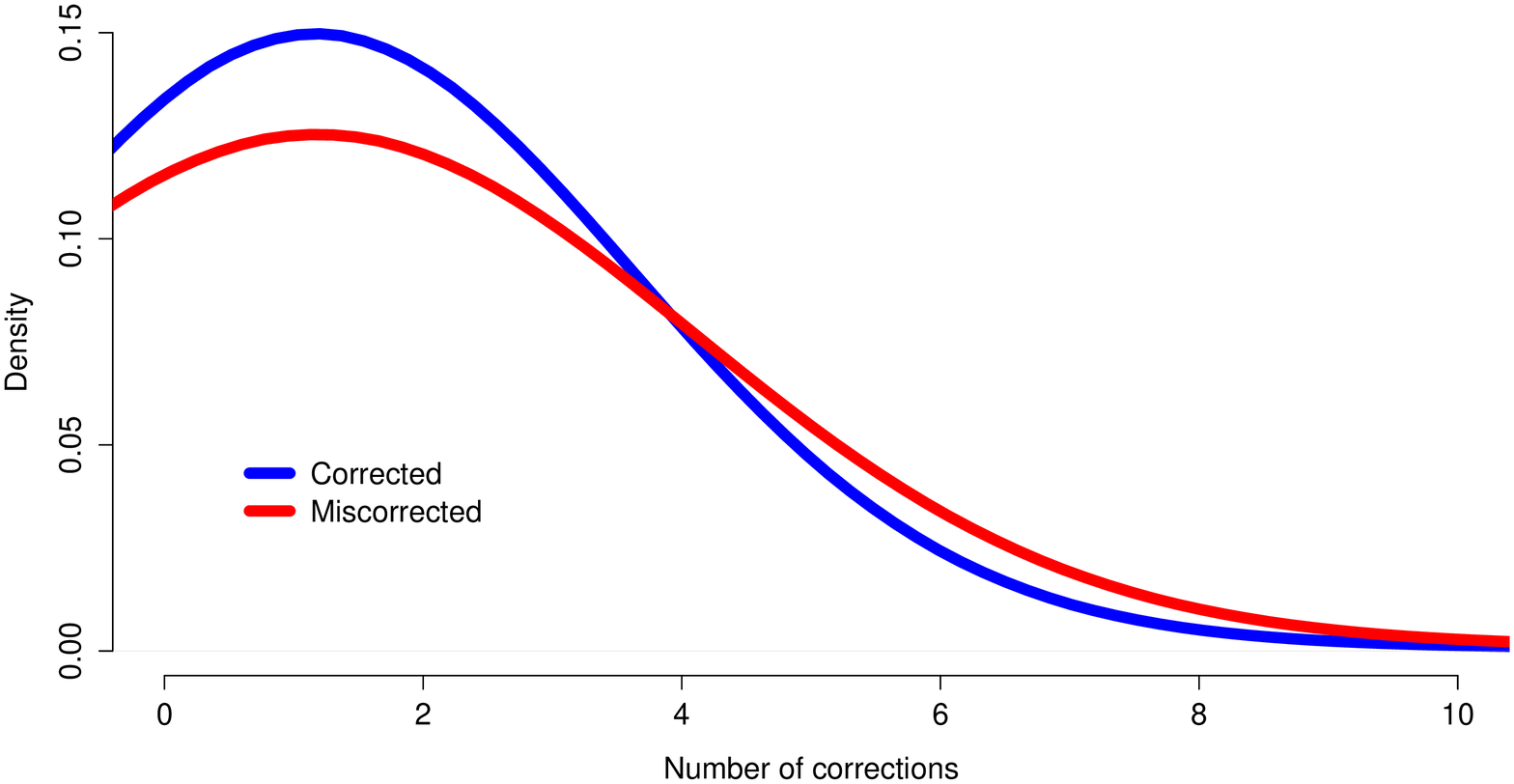}}

\noindent
\begin{changemargin}{2cm}{2cm}
Fig. 5. Reads miscorrected by \textsc{Reptile} have more corrections, on average, than to appropriately corrected reads.
\end{changemargin}
\vspace{10mm}

\noindent
Though sequence read error correction failed to have a large effect on global assembly metrics, there was substantial improvement at the nucleotide level.  Indeed, these more fine scaled effects are both harder to assay, particularly in non-model organisms, and also potentially more damaging. For instance, one popular application for transcriptome assembly is population genomics.  Most population genomics analysis are fundamentally based on estimates of polymorphism, and higher polymorphism, stemming from error, may bias results in unpredictable ways. In addition to error's effects on estimation of polymorphism, researchers interested in studying functional biology may also be impacted. Here, insertion errors may create nonsensical amino acid translation of a coding sequence, while more common substitution errors may form premature stop codons.  Though errors remain even after error correction, a reduction in magnitude of error is certainly something worth pursuing.

\section*{Methods}
Because we were interested in understanding the effects of error correction on the assembly of vertebrate transcriptome assembly, we elected to use coding sequences greater than 200nt in length from the \textit{Mus musculus} reference genome (GRCm37.71), available at \url{http://uswest.ensembl.org/Mus_musculus/Info/Index}.  Thirty million 100nt paired-end Illumina reads were simulated with the program \textsc{Flux Simulator} \citep{Griebel:2012ti} which attempts to simulate a realistic Illumina RNAseq dataset, incorporating biases related to library construction and sequencing. Thirty million PE reads were simulated as this sequencing effort was suggested to be optimal for studies of whole-animal non-model transcriptomes \citep{Francis:2013gc}. Sequencing error increased along the length of the read, as per program default. Patterns of gene expression were modeled to follow patterns typically seen in studies of Eukaryotic gene expression. The \textsc{Flux Simulator} requires the use of a parameter file, which is available at \url{https://gist.github.com/macmanes/5859902}.  \\

\noindent
In addition to analyses conducted on a simulated dataset, we used the well-characterized mouse dataset included with the Trinity software package (\url{http://sourceforge.net/projects/trinityrnaseq/files/misc/MouseRNASEQ/mouse_SS_rnaseq.50M.fastqs.tgz/download}) to validate the observed patterns using an empirically derived dataset.  To enable comparison between the simulated and empiric dataset, we randomly selected a subset of this dataset consisting of 30 million PE reads. \\

\noindent
Quality metrics for simulated and experimental raw reads were generated using the program \textsc{SolexaQA} \citep{Cox:2010ch}, and visualized using R \citep{RALanguageandEn:wf}. Patterns of gene expression were validated using the software packages \textsc{Bowtie2} \citep{Trapnell:2010kd} and \textsc{eXpress} \citep{Roberts:2012dh}. All computational work was performed on a 16-core 36GB RAM Linux Ubuntu workstation. \\

\noindent
Error correction was performed on both simulated and empirical datasets using four different error correction software packages. These included \textsc{Seecer}, \textsc{AllPathsLG} error correction, \textsc{Reptile}, and \textsc{Sga}. These specific methods were chosen in an attempt to cover the breadth of analytical methods currently used for error correction. Indeed, each of these programs implements a different computational strategy for error correction, and therefore their success, and ultimate effects on assembly accuracy are expected to vary.  In addition, several of these packages have been included in a recent review of error correction methods, with one of these (\textsc{Reptile}) having been shown to be amongst the most accurate \citep{Yang:2013ck}.  \\

\noindent
Though error correction has been a part of the \textsc{AllPathsLG} genome assembler for the past several versions, only recently has a stand-along version of their python-based error correction module (\url{http://www.broadinstitute.org/software/allpaths-lg/blog/?p=577}), which leverages several of the AllPaths subroutines, become available.  With exception to the minimum kmer frequency, which was set to 0 (unique kmers retained in the final corrected dataset), the \textsc{AllPathsLG} error correction software was run using default settings for correcting errors contained within the raw sequencing reads.  Code for running the program is available at \url{https://gist.github.com/macmanes/5859931}.   \\

\noindent
Error correction using the software package \textsc{Reptile} requires the optimization of several parameters via an included set of scripts, and therefore several runs of the program. To correct errors contained within the raw dataset, we set kmer size to 25 (\textit{KmerLen=25}), and the maximum error rate to 2\% (\textit{MaxErrRare=0.02}). Kmer=25 was selected to most closely match the kmer size used by the assembler \textsc{Trinity}. We empirically determined optimal values for \textit{T\_expGoodCnt} and \textit{T\_card} using multiple independent program executions.  \textsc{Reptile} requires the use of a parameter file, which is available at \url{https://gist.github.com/macmanes/5859947}.   \\

\noindent
The software package \textsc{SGA} was also used to correct simulated and empiric Illumina reads. This program, like \textsc{AllPaths-LG}, allows its error correction module to be applied independent of the rest of the pipeline. These preliminary steps, preprocessing, indexing, and error correction were run with default settings, with exception to the kmer size, which was set to 25.  \\

\noindent
Lastly, the software package \textsc{Seecer} was used to error correct the raw read dataset.  The software package is fundamentally different than the other packages, in that it was designed for with RNAseq reads in mind. We ran \textsc{Seecer} using default settings.  \\
 
\noindent
Transcriptome assemblies were generated using the default settings of the program \textsc{Trinity} \citep{Grabherr:2011jb}.  Code for running \textsc{Trinity} is available at \url{https://gist.github.com/macmanes/5859956}. Assemblies were evaluated using a variety of different metrics. First, \textsc{Blast+} \citep{Camacho:2009fc} was used to match assembled transcripts to their reference.  \textsc{TransDecoder} (\url{http://transdecoder.sourceforge.net/}) was used to identify full-length transcripts. For analysis of nucleotide mismatch, we elected to analyze a 'high-confidence' portion of out dataset as multiple hits and low quality BLAT matches could significantly bias results. To subset the data, we chose to include only contigs whose identity was $\geq$ 99\% similar to, and covering at least 90\% of the reference sequence.  The program \textsc{Blat} \citep{Kent:2002jd} was used to identify and count nucleotide mismatches between reconstructed transcripts in the high-confidence datasets and their corresponding reference.  Differences were visualized using the program R. \\

\subsection*{Conclusions}
To evaluate the effects of correction of sequencing error on assembly accuracy, we generated a simulated Illumina dataset, which consisted of 30M paired-end reads.  In addition, we applied the selected error correction strategy to an empirically derived \textit{Mus musculus} dataset. We attempted error correction using four popular error correction software packages, and evaluated their effect on assembly.  Though originally developed with genome sequencing in mind, we found that all tested methods do correct mRNAseq reads, and increase assembly accuracy, though \textsc{Reptile} appeared to have the most favorable effect. This study demonstrates the utility of error correction, and proposes that it become a routine step in the processing of Illumina sequence data. \\

\section*{Acknowledgments}
This paper was greatly improved by suggestions from members of the Eisen Lab, and from two named reviewers, C. Titus Brown and Mick Watson.  

\singlespacing
\bibliography{formatted.bib}

\begin{thebibliography}{40}
\expandafter\ifx\csname natexlab\endcsname\relax\def\natexlab#1{#1}\fi
\providecommand{\url}[1]{\texttt{#1}}
\providecommand{\href}[2]{#2}
\providecommand{\path}[1]{#1}
\providecommand{\DOIprefix}{doi:}
\providecommand{\ArXivprefix}{arXiv:}
\providecommand{\URLprefix}{URL: }
\providecommand{\Pubmedprefix}{pmid:}
\providecommand{\doi}[1]{\href{http://dx.doi.org/#1}{\path{#1}}}
\providecommand{\Pubmed}[1]{\href{pmid:#1}{\path{#1}}}
\providecommand{\bibinfo}[2]{#2}
\ifx\xfnm\relax \def\xfnm[#1]{\unskip,\space#1}\fi
\bibitem[{Auer and Doerge(2011)}]{Auer:2011bt}
\bibinfo{author}{Auer, P.L.}, \bibinfo{author}{Doerge, R.W.},
  \bibinfo{year}{2011}.
\newblock \bibinfo{title}{{A Two-Stage Poisson Model for Testing RNA-Seq
  Data}}.
\newblock \bibinfo{journal}{Statistical Applications in Genetics and Molecular
  Biology} \bibinfo{volume}{10}, \bibinfo{pages}{1--26}.
\bibitem[{Bradnam et~al.(2013)Bradnam, Fass, Alexandrov, Baranay, Bechner,
  Boisvert10, Chapman, Chapuis, Chikhi, Chitsaz, Chou, Corbeil, Del~Fabbro,
  Docking, Durbin, Earl, Emrich, Fedotov, Fonseca, Ganapathy, Gibbs, Gnerre,
  Godzaridis, Goldstein, Haimel, Hall, Haussler, Hiatt, Ho, Howard, Hunt,
  Jackman, Jaffe, Jarvis, Jiang, Kazakov, Kersey, Kitzman, Knight, Koren, Lam,
  Lavenier, Laviolette, Li, Li, Liu, Liu, Luo, MacCallum, MacManes, Maillet,
  Melnikov, Vieira, Naquin, Ning, Otto, Paten, Paulo, Phillippy, Pina-Martins,
  Place, Przybylski, Qin, Qu, Ribeiro, Richards, Rokhsar, Ruby, Scalabrin,
  Schatz, Schwartz, Sergushichev, Sharpe, Shaw, Shendure, Shi, Simpson, Song,
  Tsarev, Vezzi, Vicedomini, Wang, Worley, Yin, Yiu, Yuan, Zhang, Zhang, Zhou
  and Korf}]{Bradnam:2013uu}
\bibinfo{author}{Bradnam, K.R.}, \bibinfo{author}{Fass, J.N.},
  \bibinfo{author}{Alexandrov, A.}, \bibinfo{author}{Baranay, P.},
  \bibinfo{author}{Bechner, M.}, \bibinfo{author}{Boisvert10, S.},
  \bibinfo{author}{Chapman, J.A.}, \bibinfo{author}{Chapuis, G.},
  \bibinfo{author}{Chikhi, R.}, \bibinfo{author}{Chitsaz, H.},
  \bibinfo{author}{Chou, W.C.}, \bibinfo{author}{Corbeil, J.},
  \bibinfo{author}{Del~Fabbro, C.}, \bibinfo{author}{Docking, T.R.},
  \bibinfo{author}{Durbin, R.}, \bibinfo{author}{Earl, D.},
  \bibinfo{author}{Emrich, S.}, \bibinfo{author}{Fedotov, P.},
  \bibinfo{author}{Fonseca, N.A.}, \bibinfo{author}{Ganapathy, G.},
  \bibinfo{author}{Gibbs, R.A.}, \bibinfo{author}{Gnerre, S.},
  \bibinfo{author}{Godzaridis, {\'E}.}, \bibinfo{author}{Goldstein, S.},
  \bibinfo{author}{Haimel, M.}, \bibinfo{author}{Hall, G.},
  \bibinfo{author}{Haussler, D.}, \bibinfo{author}{Hiatt, J.B.},
  \bibinfo{author}{Ho, I.Y.}, \bibinfo{author}{Howard, J.},
  \bibinfo{author}{Hunt, M.}, \bibinfo{author}{Jackman, S.D.},
  \bibinfo{author}{Jaffe, D.B.}, \bibinfo{author}{Jarvis, E.},
  \bibinfo{author}{Jiang, H.}, \bibinfo{author}{Kazakov, S.},
  \bibinfo{author}{Kersey, P.J.}, \bibinfo{author}{Kitzman, J.O.},
  \bibinfo{author}{Knight, J.R.}, \bibinfo{author}{Koren, S.},
  \bibinfo{author}{Lam, T.W.}, \bibinfo{author}{Lavenier, D.},
  \bibinfo{author}{Laviolette, F.}, \bibinfo{author}{Li, Y.},
  \bibinfo{author}{Li, Z.}, \bibinfo{author}{Liu, B.}, \bibinfo{author}{Liu,
  Y.}, \bibinfo{author}{Luo, R.}, \bibinfo{author}{MacCallum, I.},
  \bibinfo{author}{MacManes, M.D.}, \bibinfo{author}{Maillet, N.},
  \bibinfo{author}{Melnikov, S.}, \bibinfo{author}{Vieira, B.M.},
  \bibinfo{author}{Naquin, D.}, \bibinfo{author}{Ning, Z.},
  \bibinfo{author}{Otto, T.D.}, \bibinfo{author}{Paten, B.},
  \bibinfo{author}{Paulo, O.S.}, \bibinfo{author}{Phillippy, A.M.},
  \bibinfo{author}{Pina-Martins, F.}, \bibinfo{author}{Place, M.},
  \bibinfo{author}{Przybylski, D.}, \bibinfo{author}{Qin, X.},
  \bibinfo{author}{Qu, C.}, \bibinfo{author}{Ribeiro, F.J.},
  \bibinfo{author}{Richards, S.}, \bibinfo{author}{Rokhsar, D.S.},
  \bibinfo{author}{Ruby, J.G.}, \bibinfo{author}{Scalabrin, S.},
  \bibinfo{author}{Schatz, M.C.}, \bibinfo{author}{Schwartz, D.C.},
  \bibinfo{author}{Sergushichev, A.}, \bibinfo{author}{Sharpe, T.},
  \bibinfo{author}{Shaw, T.I.}, \bibinfo{author}{Shendure, J.},
  \bibinfo{author}{Shi, Y.}, \bibinfo{author}{Simpson, J.T.},
  \bibinfo{author}{Song, H.}, \bibinfo{author}{Tsarev, F.},
  \bibinfo{author}{Vezzi, F.}, \bibinfo{author}{Vicedomini, R.},
  \bibinfo{author}{Wang, J.}, \bibinfo{author}{Worley, K.C.},
  \bibinfo{author}{Yin, S.}, \bibinfo{author}{Yiu, S.M.},
  \bibinfo{author}{Yuan, J.}, \bibinfo{author}{Zhang, G.},
  \bibinfo{author}{Zhang, H.}, \bibinfo{author}{Zhou, S.},
  \bibinfo{author}{Korf, I.F.}, \bibinfo{year}{2013}.
\newblock \bibinfo{title}{{Assemblathon 2: evaluating \textit{de novo} methods
  of genome assembly in three vertebrate species}}.
\newblock \bibinfo{journal}{arXiv.org}
  \href{http://arxiv.org/abs/1301.5406v1}{\tt arXiv:1301.5406v1}.
\bibitem[{Camacho et~al.(2009)Camacho, Coulouris, Avagyan, Ma, Papadopoulos,
  Bealer and Madden}]{Camacho:2009fc}
\bibinfo{author}{Camacho, C.}, \bibinfo{author}{Coulouris, G.},
  \bibinfo{author}{Avagyan, V.}, \bibinfo{author}{Ma, N.},
  \bibinfo{author}{Papadopoulos, J.}, \bibinfo{author}{Bealer, K.},
  \bibinfo{author}{Madden, T.L.}, \bibinfo{year}{2009}.
\newblock \bibinfo{title}{{BLAST+: architecture and applications}}.
\newblock \bibinfo{journal}{BMC Bioinformatics} \bibinfo{volume}{10},
  \bibinfo{pages}{421}.
\bibitem[{Chaisson et~al.(2004)Chaisson, Pevzner and Tang}]{Chaisson:2004kf}
\bibinfo{author}{Chaisson, M.}, \bibinfo{author}{Pevzner, P.},
  \bibinfo{author}{Tang, H.X.}, \bibinfo{year}{2004}.
\newblock \bibinfo{title}{{Fragment assembly with short reads}}.
\newblock \bibinfo{journal}{Bioinformatics (Oxford, England)}
  \bibinfo{volume}{20}, \bibinfo{pages}{2067--2074}.
\bibitem[{Chen et~al.(2011)Chen, Liu, Ng, Nadarajah, Kaufman, Yang and
  Deng}]{Chen:2011ba}
\bibinfo{author}{Chen, Z.}, \bibinfo{author}{Liu, J.}, \bibinfo{author}{Ng,
  H.K.T.}, \bibinfo{author}{Nadarajah, S.}, \bibinfo{author}{Kaufman, H.L.},
  \bibinfo{author}{Yang, J.Y.}, \bibinfo{author}{Deng, Y.},
  \bibinfo{year}{2011}.
\newblock \bibinfo{title}{{Statistical methods on detecting differentially
  expressed genes for RNA-seq data}}.
\newblock \bibinfo{journal}{BMC Systems Biology} \bibinfo{volume}{5},
  \bibinfo{pages}{S1}.
\bibitem[{Conway and Bromage(2011)}]{Conway:2011iv}
\bibinfo{author}{Conway, T.C.}, \bibinfo{author}{Bromage, a.J.},
  \bibinfo{year}{2011}.
\newblock \bibinfo{title}{{Succinct data structures for assembling large
  genomes.}}
\newblock \bibinfo{journal}{Bioinformatics (Oxford, England)}
  \bibinfo{volume}{27}, \bibinfo{pages}{479--486}.
\bibitem[{Cox et~al.(2010)Cox, Peterson and Biggs}]{Cox:2010ch}
\bibinfo{author}{Cox, M.P.}, \bibinfo{author}{Peterson, D.A.},
  \bibinfo{author}{Biggs, P.J.}, \bibinfo{year}{2010}.
\newblock \bibinfo{title}{{SolexaQA: At-a-glance quality assessment of Illumina
  second-generation sequencing data}}.
\newblock \bibinfo{journal}{BMC Bioinformatics} \bibinfo{volume}{11},
  \bibinfo{pages}{485}.
\bibitem[{Earl et~al.(2011)Earl, Bradnam, St~John, Darling, Lin, Fass, Yu,
  Buffalo, Zerbino, Diekhans, Nguyen, Ariyaratne, Sung, Ning, Haimel, Simpson,
  Fonseca, Birol, Docking, Ho, Rokhsar, Chikhi, Lavenier, Chapuis, Naquin,
  Maillet, Schatz, Kelley, Phillippy, Koren, Yang, Wu, Chou, Srivastava, Shaw,
  Ruby, Skewes-Cox, Betegon, Dimon, Solovyev, Seledtsov, Kosarev, Vorobyev,
  Ramirez-Gonzalez, Leggett, MacLean, Xia, Luo, Li, Xie, Liu, Gnerre,
  Maccallum, Przybylski, Ribeiro, Yin, Sharpe, Hall, Kersey, Durbin, Jackman,
  Chapman, Huang, DeRisi, Caccamo, Li, Jaffe, Green, Haussler, Korf and
  Paten}]{Earl:2011gt}
\bibinfo{author}{Earl, D.}, \bibinfo{author}{Bradnam, K.},
  \bibinfo{author}{St~John, J.}, \bibinfo{author}{Darling, A.},
  \bibinfo{author}{Lin, D.}, \bibinfo{author}{Fass, J.}, \bibinfo{author}{Yu,
  H.O.K.}, \bibinfo{author}{Buffalo, V.}, \bibinfo{author}{Zerbino, D.R.},
  \bibinfo{author}{Diekhans, M.}, \bibinfo{author}{Nguyen, N.},
  \bibinfo{author}{Ariyaratne, P.N.}, \bibinfo{author}{Sung, W.K.},
  \bibinfo{author}{Ning, Z.}, \bibinfo{author}{Haimel, M.},
  \bibinfo{author}{Simpson, J.T.}, \bibinfo{author}{Fonseca, N.A.},
  \bibinfo{author}{Birol, I.}, \bibinfo{author}{Docking, T.R.},
  \bibinfo{author}{Ho, I.Y.}, \bibinfo{author}{Rokhsar, D.S.},
  \bibinfo{author}{Chikhi, R.}, \bibinfo{author}{Lavenier, D.},
  \bibinfo{author}{Chapuis, G.}, \bibinfo{author}{Naquin, D.},
  \bibinfo{author}{Maillet, N.}, \bibinfo{author}{Schatz, M.C.},
  \bibinfo{author}{Kelley, D.R.}, \bibinfo{author}{Phillippy, A.M.},
  \bibinfo{author}{Koren, S.}, \bibinfo{author}{Yang, S.P.},
  \bibinfo{author}{Wu, W.}, \bibinfo{author}{Chou, W.C.},
  \bibinfo{author}{Srivastava, A.}, \bibinfo{author}{Shaw, T.I.},
  \bibinfo{author}{Ruby, J.G.}, \bibinfo{author}{Skewes-Cox, P.},
  \bibinfo{author}{Betegon, M.}, \bibinfo{author}{Dimon, M.T.},
  \bibinfo{author}{Solovyev, V.}, \bibinfo{author}{Seledtsov, I.},
  \bibinfo{author}{Kosarev, P.}, \bibinfo{author}{Vorobyev, D.},
  \bibinfo{author}{Ramirez-Gonzalez, R.}, \bibinfo{author}{Leggett, R.},
  \bibinfo{author}{MacLean, D.}, \bibinfo{author}{Xia, F.},
  \bibinfo{author}{Luo, R.}, \bibinfo{author}{Li, Z.}, \bibinfo{author}{Xie,
  Y.}, \bibinfo{author}{Liu, B.}, \bibinfo{author}{Gnerre, S.},
  \bibinfo{author}{Maccallum, I.}, \bibinfo{author}{Przybylski, D.},
  \bibinfo{author}{Ribeiro, F.J.}, \bibinfo{author}{Yin, S.},
  \bibinfo{author}{Sharpe, T.}, \bibinfo{author}{Hall, G.},
  \bibinfo{author}{Kersey, P.J.}, \bibinfo{author}{Durbin, R.},
  \bibinfo{author}{Jackman, S.D.}, \bibinfo{author}{Chapman, J.A.},
  \bibinfo{author}{Huang, X.}, \bibinfo{author}{DeRisi, J.L.},
  \bibinfo{author}{Caccamo, M.}, \bibinfo{author}{Li, Y.},
  \bibinfo{author}{Jaffe, D.B.}, \bibinfo{author}{Green, R.E.},
  \bibinfo{author}{Haussler, D.}, \bibinfo{author}{Korf, I.},
  \bibinfo{author}{Paten, B.}, \bibinfo{year}{2011}.
\newblock \bibinfo{title}{{Assemblathon 1: A competitive assessment of
  \textit{de novo} short read assembly methods}}.
\newblock \bibinfo{journal}{Genome Research} \bibinfo{volume}{21},
  \bibinfo{pages}{2224--2241}.
\bibitem[{Francis et~al.(2013)Francis, Christianson, Kiko, Powers, Shaner and
  Haddock}]{Francis:2013gc}
\bibinfo{author}{Francis, W.R.}, \bibinfo{author}{Christianson, L.M.},
  \bibinfo{author}{Kiko, R.}, \bibinfo{author}{Powers, M.L.},
  \bibinfo{author}{Shaner, N.C.}, \bibinfo{author}{Haddock, S.H.D.},
  \bibinfo{year}{2013}.
\newblock \bibinfo{title}{{A comparison across non-model animals suggests an
  optimal sequencing depth for \textit{de novo} transcriptome assembly}}.
\newblock \bibinfo{journal}{BMC Genomics} \bibinfo{volume}{14},
  \bibinfo{pages}{167}.
\bibitem[{Gnerre et~al.(2011)Gnerre, MacCallum, Przybylski, Ribeiro, Burton,
  Walker, Sharpe, Hall, Shea, Sykes, Berlin, Aird, Costello, Daza, Williams,
  Nicol, Gnirke, Nusbaum, Lander and Jaffe}]{Gnerre:2011fd}
\bibinfo{author}{Gnerre, S.}, \bibinfo{author}{MacCallum, I.},
  \bibinfo{author}{Przybylski, D.}, \bibinfo{author}{Ribeiro, F.J.},
  \bibinfo{author}{Burton, J.N.}, \bibinfo{author}{Walker, B.J.},
  \bibinfo{author}{Sharpe, T.}, \bibinfo{author}{Hall, G.},
  \bibinfo{author}{Shea, T.P.}, \bibinfo{author}{Sykes, S.},
  \bibinfo{author}{Berlin, A.M.}, \bibinfo{author}{Aird, D.},
  \bibinfo{author}{Costello, M.}, \bibinfo{author}{Daza, R.},
  \bibinfo{author}{Williams, L.}, \bibinfo{author}{Nicol, R.},
  \bibinfo{author}{Gnirke, a.}, \bibinfo{author}{Nusbaum, C.},
  \bibinfo{author}{Lander, E.S.}, \bibinfo{author}{Jaffe, D.B.},
  \bibinfo{year}{2011}.
\newblock \bibinfo{title}{{High-quality draft assemblies of mammalian genomes
  from massively parallel sequence data.}}
\newblock \bibinfo{journal}{Proceedings of the National Academy of Sciences}
  \bibinfo{volume}{108}, \bibinfo{pages}{1513--1518}.
\bibitem[{Grabherr et~al.(2011)Grabherr, Haas, Yassour, Levin, Thompson, Amit,
  Adiconis, Fan, Raychowdhury, Zeng, Chen, Mauceli, Hacohen, Gnirke, Rhind,
  di~Palma, Birren, Nusbaum, Lindblad-Toh, Friedman and
  Regev}]{Grabherr:2011jb}
\bibinfo{author}{Grabherr, M.G.}, \bibinfo{author}{Haas, B.J.},
  \bibinfo{author}{Yassour, M.}, \bibinfo{author}{Levin, J.Z.},
  \bibinfo{author}{Thompson, D.A.}, \bibinfo{author}{Amit, I.},
  \bibinfo{author}{Adiconis, X.}, \bibinfo{author}{Fan, L.},
  \bibinfo{author}{Raychowdhury, R.}, \bibinfo{author}{Zeng, Q.},
  \bibinfo{author}{Chen, Z.}, \bibinfo{author}{Mauceli, E.},
  \bibinfo{author}{Hacohen, N.}, \bibinfo{author}{Gnirke, a.},
  \bibinfo{author}{Rhind, N.}, \bibinfo{author}{di~Palma, F.},
  \bibinfo{author}{Birren, B.W.}, \bibinfo{author}{Nusbaum, C.},
  \bibinfo{author}{Lindblad-Toh, K.}, \bibinfo{author}{Friedman, N.},
  \bibinfo{author}{Regev, A.}, \bibinfo{year}{2011}.
\newblock \bibinfo{title}{{Full-length transcriptome assembly from RNA-Seq data
  without a reference genome.}}
\newblock \bibinfo{journal}{Nature Biotechnology} \bibinfo{volume}{29},
  \bibinfo{pages}{644--652}.
\bibitem[{Griebel et~al.(2012)Griebel, Zacher, Ribeca, Raineri, Lacroix,
  Guig{\'o} and Sammeth}]{Griebel:2012ti}
\bibinfo{author}{Griebel, T.}, \bibinfo{author}{Zacher, B.},
  \bibinfo{author}{Ribeca, P.}, \bibinfo{author}{Raineri, E.},
  \bibinfo{author}{Lacroix, V.}, \bibinfo{author}{Guig{\'o}, R.},
  \bibinfo{author}{Sammeth, M.}, \bibinfo{year}{2012}.
\newblock \bibinfo{title}{{Modelling and simulating generic RNA-Seq experiments
  with the flux simulator.}}
\newblock \bibinfo{journal}{Nucleic Acids Research} \bibinfo{volume}{40},
  \bibinfo{pages}{10073--10083}.
\bibitem[{Haas et~al.(2013)Haas, Papanicolaou, Yassour, Grabherr, Blood,
  Bowden, Couger, Eccles, Li, Lieber, MacManes, Ott, Orvis, Pochet, Strozzi,
  Weeks, Westerman, William, Dewey, Henschel, LeDuc, Friedman and
  Regev}]{Haas:jq}
\bibinfo{author}{Haas, B.J.}, \bibinfo{author}{Papanicolaou, A.},
  \bibinfo{author}{Yassour, M.}, \bibinfo{author}{Grabherr, M.},
  \bibinfo{author}{Blood, P.}, \bibinfo{author}{Bowden, J.},
  \bibinfo{author}{Couger, M.}, \bibinfo{author}{Eccles, D.},
  \bibinfo{author}{Li, B.}, \bibinfo{author}{Lieber, M.},
  \bibinfo{author}{MacManes, M.D.}, \bibinfo{author}{Ott, M.},
  \bibinfo{author}{Orvis, J.}, \bibinfo{author}{Pochet, N.},
  \bibinfo{author}{Strozzi, F.}, \bibinfo{author}{Weeks, N.},
  \bibinfo{author}{Westerman, R.}, \bibinfo{author}{William, T.},
  \bibinfo{author}{Dewey, C.N.}, \bibinfo{author}{Henschel, R.},
  \bibinfo{author}{LeDuc, R.G.}, \bibinfo{author}{Friedman, N.},
  \bibinfo{author}{Regev, A.}, \bibinfo{year}{2013}.
\newblock \bibinfo{title}{{\emph{\textit{De novo}} transcript sequence
  reconstruction from RNA-seq using the Trinity platform for reference
  generation and analysis}}.
\newblock \bibinfo{journal}{Nature protocols} , \bibinfo{pages}{1--21}.
\bibitem[{Hsu et~al.(2012)Hsu, Chien, Hu, Chen, Wu, Feng, Haymer and
  Chen}]{Hsu:2012dg}
\bibinfo{author}{Hsu, J.C.}, \bibinfo{author}{Chien, T.Y.},
  \bibinfo{author}{Hu, C.C.}, \bibinfo{author}{Chen, M.J.M.},
  \bibinfo{author}{Wu, W.J.}, \bibinfo{author}{Feng, H.T.},
  \bibinfo{author}{Haymer, D.S.}, \bibinfo{author}{Chen, C.Y.},
  \bibinfo{year}{2012}.
\newblock \bibinfo{title}{{Discovery of Genes Related to Insecticide Resistance
  in \emph{Bactrocera dorsalis} by Functional Genomic Analysis of a \emph{De
  Novo} Assembled Transcriptome}}.
\newblock \bibinfo{journal}{PLOS ONE} \bibinfo{volume}{7},
  \bibinfo{pages}{e40950}.
\bibitem[{Hu et~al.(2011)Hu, Zhu, Taylor, Liu and Qin}]{Hu:2011cn}
\bibinfo{author}{Hu, M.}, \bibinfo{author}{Zhu, Y.}, \bibinfo{author}{Taylor,
  J.M.G.}, \bibinfo{author}{Liu, J.S.}, \bibinfo{author}{Qin, Z.S.},
  \bibinfo{year}{2011}.
\newblock \bibinfo{title}{{Using Poisson mixed-effects model to quantify
  transcript-level gene expression in RNA-Seq}}.
\newblock \bibinfo{journal}{Bioinformatics (Oxford, England)}
  \bibinfo{volume}{28}, \bibinfo{pages}{63--68}.
\bibitem[{Jiang and Wong(2009)}]{Jiang:2009bw}
\bibinfo{author}{Jiang, H.}, \bibinfo{author}{Wong, W.H.},
  \bibinfo{year}{2009}.
\newblock \bibinfo{title}{{Statistical inferences for isoform expression in
  RNA-Seq}}.
\newblock \bibinfo{journal}{Bioinformatics (Oxford, England)}
  \bibinfo{volume}{25}, \bibinfo{pages}{1026--1032}.
\bibitem[{Kao et~al.(2011)Kao, Chan and Song}]{Kao:2011bx}
\bibinfo{author}{Kao, W.C.}, \bibinfo{author}{Chan, a.H.},
  \bibinfo{author}{Song, Y.S.}, \bibinfo{year}{2011}.
\newblock \bibinfo{title}{{ECHO: a reference-free short-read error correction
  algorithm.}}
\newblock \bibinfo{journal}{Genome Research} \bibinfo{volume}{21},
  \bibinfo{pages}{1181--1192}.
\bibitem[{Kelley et~al.(2010)Kelley, Schatz and Salzberg}]{Kelley:2010kg}
\bibinfo{author}{Kelley, D.R.}, \bibinfo{author}{Schatz, M.C.},
  \bibinfo{author}{Salzberg, S.L.}, \bibinfo{year}{2010}.
\newblock \bibinfo{title}{{Quake: quality-aware detection and correction of
  sequencing errors}}.
\newblock \bibinfo{journal}{Genome Biology} \bibinfo{volume}{11},
  \bibinfo{pages}{R116}.
\bibitem[{Kent(2002)}]{Kent:2002jd}
\bibinfo{author}{Kent, W.J.}, \bibinfo{year}{2002}.
\newblock \bibinfo{title}{{BLAT---The BLAST-Like Alignment Tool}}.
\newblock \bibinfo{journal}{Genome Research} \bibinfo{volume}{12},
  \bibinfo{pages}{656--664}.
\bibitem[{Le et~al.(2013)Le, Schulz, McCauley, Hinman and
  Bar-Joseph}]{Le:2013dy}
\bibinfo{author}{Le, H.S.}, \bibinfo{author}{Schulz, M.H.},
  \bibinfo{author}{McCauley, B.M.}, \bibinfo{author}{Hinman, V.F.},
  \bibinfo{author}{Bar-Joseph, Z.}, \bibinfo{year}{2013}.
\newblock \bibinfo{title}{{Probabilistic error correction for RNA sequencing}}.
\newblock \bibinfo{journal}{Nucleic Acids Research} .
\bibitem[{Linnen et~al.(2013)Linnen, Poh, Peterson, Barrett, Larson, Jensen and
  Hoekstra}]{Linnen:2013ir}
\bibinfo{author}{Linnen, C.R.}, \bibinfo{author}{Poh, Y.P.},
  \bibinfo{author}{Peterson, B.K.}, \bibinfo{author}{Barrett, R.D.H.},
  \bibinfo{author}{Larson, J.G.}, \bibinfo{author}{Jensen, J.D.},
  \bibinfo{author}{Hoekstra, H.E.}, \bibinfo{year}{2013}.
\newblock \bibinfo{title}{{Adaptive Evolution of Multiple Traits Through
  Multiple Mutations at a Single Gene}}.
\newblock \bibinfo{journal}{Science (New York, NY)} \bibinfo{volume}{339},
  \bibinfo{pages}{1312--1316}.
\bibitem[{Liu et~al.(2012)Liu, Yuan, Yiu, Li, Xie, Chen, Shi, Zhang, Li, Lam
  and Luo}]{Liu:2012iv}
\bibinfo{author}{Liu, B.}, \bibinfo{author}{Yuan, J.}, \bibinfo{author}{Yiu,
  S.M.}, \bibinfo{author}{Li, Z.}, \bibinfo{author}{Xie, Y.},
  \bibinfo{author}{Chen, Y.}, \bibinfo{author}{Shi, Y.},
  \bibinfo{author}{Zhang, H.}, \bibinfo{author}{Li, Y.}, \bibinfo{author}{Lam,
  T.W.}, \bibinfo{author}{Luo, R.}, \bibinfo{year}{2012}.
\newblock \bibinfo{title}{{COPE: an accurate k-mer-based pair-end reads
  connection tool to facilitate genome assembly}}.
\newblock \bibinfo{journal}{Bioinformatics (Oxford, England)}
  \bibinfo{volume}{28}, \bibinfo{pages}{2870--2874}.
\bibitem[{Liu et~al.(2011)Liu, Schmidt and Maskell}]{Liu:2011ez}
\bibinfo{author}{Liu, Y.}, \bibinfo{author}{Schmidt, B.},
  \bibinfo{author}{Maskell, D.L.}, \bibinfo{year}{2011}.
\newblock \bibinfo{title}{{Parallelized short read assembly of large genomes
  using \textit{de Bruijn} graphs.}}
\newblock \bibinfo{journal}{BMC Bioinformatics} \bibinfo{volume}{12},
  \bibinfo{pages}{354}.
\bibitem[{MacManes and Lacey(2012)}]{MacManes:2012bu}
\bibinfo{author}{MacManes, M.D.}, \bibinfo{author}{Lacey, E.A.},
  \bibinfo{year}{2012}.
\newblock \bibinfo{title}{{The Social Brain: Transcriptome Assembly and
  Characterization of the Hippocampus from a Social Subterranean Rodent, the
  Colonial Tuco-Tuco (\textit{Ctenomys} \textit{sociabilis})}}.
\newblock \bibinfo{journal}{PLOS ONE} \bibinfo{volume}{7},
  \bibinfo{pages}{e45524}.
\bibitem[{Marioni et~al.(2008)Marioni, Mason, Mane, Stephens and
  Gilad}]{Marioni:2008bg}
\bibinfo{author}{Marioni, J.C.}, \bibinfo{author}{Mason, C.E.},
  \bibinfo{author}{Mane, S.M.}, \bibinfo{author}{Stephens, M.},
  \bibinfo{author}{Gilad, Y.}, \bibinfo{year}{2008}.
\newblock \bibinfo{title}{{RNA-seq: An assessment of technical reproducibility
  and comparison with gene expression arrays}}.
\newblock \bibinfo{journal}{Genome Research} \bibinfo{volume}{18},
  \bibinfo{pages}{1509--1517}.
\bibitem[{Mu\~noz Merida et~al.(2013)Mu\~noz Merida, Gonzalez-Plaza, Canada,
  Blanco, Garcia-Lopez, Rodriguez, Pedrola, Sicardo, Hernandez, De~la Rosa,
  Belaj, Gil-Borja, Luque, Martinez-Rivas, Pisano, Trelles, Valpuesta and
  Beuzon}]{MunozMerida:2013dw}
\bibinfo{author}{Mu\~noz Merida, A.}, \bibinfo{author}{Gonzalez-Plaza, J.J.},
  \bibinfo{author}{Canada, A.}, \bibinfo{author}{Blanco, A.M.},
  \bibinfo{author}{Garcia-Lopez, M.d.C.}, \bibinfo{author}{Rodriguez, J.M.},
  \bibinfo{author}{Pedrola, L.}, \bibinfo{author}{Sicardo, M.D.},
  \bibinfo{author}{Hernandez, M.L.}, \bibinfo{author}{De~la Rosa, R.},
  \bibinfo{author}{Belaj, A.}, \bibinfo{author}{Gil-Borja, M.},
  \bibinfo{author}{Luque, F.}, \bibinfo{author}{Martinez-Rivas, J.M.},
  \bibinfo{author}{Pisano, D.G.}, \bibinfo{author}{Trelles, O.},
  \bibinfo{author}{Valpuesta, V.}, \bibinfo{author}{Beuzon, C.R.},
  \bibinfo{year}{2013}.
\newblock \bibinfo{title}{{De Novo Assembly and Functional Annotation of the
  Olive (\textit{Olea europaea}) Transcriptome}}.
\newblock \bibinfo{journal}{DNA Research} \bibinfo{volume}{20},
  \bibinfo{pages}{93--108}.
\bibitem[{Miller et~al.(2010)Miller, Koren and Sutton}]{Miller:2010gxa}
\bibinfo{author}{Miller, J.R.}, \bibinfo{author}{Koren, S.},
  \bibinfo{author}{Sutton, G.}, \bibinfo{year}{2010}.
\newblock \bibinfo{title}{{Assembly algorithms for next-generation sequencing
  data}}.
\newblock \bibinfo{journal}{Genomics} \bibinfo{volume}{95},
  \bibinfo{pages}{315--327}.
\bibitem[{Mortazavi et~al.(2008)Mortazavi, Williams, Mccue, Schaeffer and
  Wold}]{Mortazavi:2008jj}
\bibinfo{author}{Mortazavi, A.}, \bibinfo{author}{Williams, B.A.},
  \bibinfo{author}{Mccue, K.}, \bibinfo{author}{Schaeffer, L.},
  \bibinfo{author}{Wold, B.}, \bibinfo{year}{2008}.
\newblock \bibinfo{title}{{Mapping and quantifying mammalian transcriptomes by
  RNA-Seq}}.
\newblock \bibinfo{journal}{Nature Methods} \bibinfo{volume}{5},
  \bibinfo{pages}{621--628}.
\bibitem[{Narum et~al.(2013)Narum, Campbell, Meyer, Miller and
  Hardy}]{Narum:2013kc}
\bibinfo{author}{Narum, S.R.}, \bibinfo{author}{Campbell, N.R.},
  \bibinfo{author}{Meyer, K.A.}, \bibinfo{author}{Miller, M.R.},
  \bibinfo{author}{Hardy, R.W.}, \bibinfo{year}{2013}.
\newblock \bibinfo{title}{{Thermal adaptation and acclimation of ectotherms
  from differing aquatic climates.}}
\newblock \bibinfo{journal}{Molecular Ecology} , \bibinfo{pages}{1--8}.
\bibitem[{Pell et~al.(2012)Pell, Hintze, Canino-Koning, Howe, Tiedje and
  Brown}]{Pell:2012id}
\bibinfo{author}{Pell, J.}, \bibinfo{author}{Hintze, A.},
  \bibinfo{author}{Canino-Koning, R.}, \bibinfo{author}{Howe, A.},
  \bibinfo{author}{Tiedje, J.M.}, \bibinfo{author}{Brown, C.T.},
  \bibinfo{year}{2012}.
\newblock \bibinfo{title}{{Scaling metagenome sequence assembly with
  probabilistic \textit{de Bruijn} graphs.}}
\newblock \bibinfo{journal}{Proceedings of the National Academy of Sciences}
  \bibinfo{volume}{109}, \bibinfo{pages}{13272--13277}.
\bibitem[{Pyrkosz et~al.(2013)Pyrkosz, Cheng and Brown}]{Pyrkosz:2013tm}
\bibinfo{author}{Pyrkosz, A.B.}, \bibinfo{author}{Cheng, H.},
  \bibinfo{author}{Brown, C.T.}, \bibinfo{year}{2013}.
\newblock \bibinfo{title}{{RNA-Seq Mapping Errors When Using Incomplete
  Reference Transcriptomes of Vertebrates}}.
\newblock \bibinfo{journal}{\url{http://arxiv.org/abs/1303.2411v1}}
  \href{http://arxiv.org/abs/1303.2411v1}{\tt arXiv:1303.2411v1}.
\bibitem[{R~Core Development~Team(2011)}]{RALanguageandEn:wf}
\bibinfo{author}{R~Core Development~Team, F.}, \bibinfo{year}{2011}.
\newblock \bibinfo{title}{{R: A Language and Environment for Statistical
  Computing}} .
\bibitem[{Roberts and Pachter(2012)}]{Roberts:2012dh}
\bibinfo{author}{Roberts, A.}, \bibinfo{author}{Pachter, L.},
  \bibinfo{year}{2012}.
\newblock \bibinfo{title}{{Streaming fragment assignment for real-time analysis
  of sequencing experiments}}.
\newblock \bibinfo{journal}{Nature Methods} , \bibinfo{pages}{1--7}.
\bibitem[{Sammeth(2009)}]{Sammeth:2009jx}
\bibinfo{author}{Sammeth, M.}, \bibinfo{year}{2009}.
\newblock \bibinfo{title}{{Complete alternative splicing events are bubbles in
  splicing graphs.}}
\newblock \bibinfo{journal}{Journal of computational biology : a journal of
  computational molecular cell biology} \bibinfo{volume}{16},
  \bibinfo{pages}{1117--1140}.
\bibitem[{Simpson and Durbin(2010)}]{Simpson:2010fd}
\bibinfo{author}{Simpson, J.T.}, \bibinfo{author}{Durbin, R.},
  \bibinfo{year}{2010}.
\newblock \bibinfo{title}{{Efficient construction of an assembly string graph
  using the FM-index}}.
\newblock \bibinfo{journal}{Bioinformatics (Oxford, England)}
  \bibinfo{volume}{26}, \bibinfo{pages}{i367--i373}.
\bibitem[{Soneson and Delorenzi(2013)}]{Soneson:2013fr}
\bibinfo{author}{Soneson, C.}, \bibinfo{author}{Delorenzi, M.},
  \bibinfo{year}{2013}.
\newblock \bibinfo{title}{{A comparison of methods for differential expression
  analysis of RNA-seq data}}.
\newblock \bibinfo{journal}{BMC Bioinformatics} \bibinfo{volume}{14},
  \bibinfo{pages}{91}.
\bibitem[{Trapnell et~al.(2010)Trapnell, Williams, Pertea, Mortazavi, Kwan, van
  Baren, Salzberg, Wold and Pachter}]{Trapnell:2010kd}
\bibinfo{author}{Trapnell, C.}, \bibinfo{author}{Williams, B.A.},
  \bibinfo{author}{Pertea, G.}, \bibinfo{author}{Mortazavi, A.},
  \bibinfo{author}{Kwan, G.}, \bibinfo{author}{van Baren, M.J.},
  \bibinfo{author}{Salzberg, S.L.}, \bibinfo{author}{Wold, B.J.},
  \bibinfo{author}{Pachter, L.}, \bibinfo{year}{2010}.
\newblock \bibinfo{title}{{Transcript assembly and quantification by RNA-Seq
  reveals unannotated transcripts and isoform switching during cell
  differentiation}}.
\newblock \bibinfo{journal}{Nature Biotechnology} \bibinfo{volume}{28},
  \bibinfo{pages}{511--U174}.
\bibitem[{Vijay et~al.(2013)Vijay, Poelstra, K{\"u}nstner and
  Wolf}]{Vijay:2012gy}
\bibinfo{author}{Vijay, N.}, \bibinfo{author}{Poelstra, J.W.},
  \bibinfo{author}{K{\"u}nstner, A.}, \bibinfo{author}{Wolf, J.B.W.},
  \bibinfo{year}{2013}.
\newblock \bibinfo{title}{{Challenges and strategies in transcriptome assembly
  and differential gene expression quantification. A comprehensive \textit{in
  silico} assessment of RNA-seq experiments.}}
\newblock \bibinfo{journal}{Molecular Ecology} \bibinfo{volume}{22},
  \bibinfo{pages}{620--634}.
\bibitem[{Yang et~al.(2013)Yang, Chockalingam and Aluru}]{Yang:2013ck}
\bibinfo{author}{Yang, X.}, \bibinfo{author}{Chockalingam, S.P.},
  \bibinfo{author}{Aluru, S.}, \bibinfo{year}{2013}.
\newblock \bibinfo{title}{{A survey of error-correction methods for
  next-generation sequencing}}.
\newblock \bibinfo{journal}{Briefings In Bioinformatics} \bibinfo{volume}{14},
  \bibinfo{pages}{56--66}.
\bibitem[{Yang et~al.(2010)Yang, Dorman and Aluru}]{Yang:2010kv}
\bibinfo{author}{Yang, X.}, \bibinfo{author}{Dorman, K.S.},
  \bibinfo{author}{Aluru, S.}, \bibinfo{year}{2010}.
\newblock \bibinfo{title}{{Reptile: representative tiling for short read error
  correction.}}
\newblock \bibinfo{journal}{Bioinformatics (Oxford, England)}
  \bibinfo{volume}{26}, \bibinfo{pages}{2526--2533}.

\end{thebibliography}
\bibliographystyle{model2-names.bst}

\end{document}